\documentclass[%
 aip,
 amsmath,amssymb,
 reprint,%
]{revtex4-1}

\usepackage{graphicx}
\usepackage{dcolumn}
\usepackage{bm}
\usepackage{xcolor}
\usepackage{booktabs}
\usepackage{mathtools, nccmath}

\usepackage[utf8]{inputenc}
\usepackage{siunitx}
\usepackage[T1]{fontenc}
\usepackage{mathptmx}
\usepackage{etoolbox}
\usepackage[normalem]{ulem}

\makeatletter
\def\@email#1#2{%
 \endgroup
 \patchcmd{\titleblock@produce}
  {\frontmatter@RRAPformat}
  {\frontmatter@RRAPformat{\produce@RRAP{*#1\href{mailto:#2}{#2}}}\frontmatter@RRAPformat}
  {}{}
}
\makeatother
\begin{document}

\preprint{AIP/123-QED}

\title[]{Thermal resistance in superconducting flip-chip assemblies}

\author{J. Hätinen*}
\author{E. Mykkänen}
\author{K. Viisanen}
\author{A. Ronzani}
\author{A. Kemppinen}
\author{L. Lehtisyrjä}
\author{J. S. Lehtinen}
\author{\\M. Prunnila}

\email{joel.hatinen@vtt.fi}

\affiliation{ 
$^{1}$VTT Technical Research Centre of Finland Ltd. Tietotie 3, Espoo, 02150, Finland
}
\date{\today}

\begin{abstract}

Cryogenic microsystems that utilize different 3D integration techniques are being actively developed, e.g., for the needs of quantum technologies. 3D integration can introduce opportunities and challenges to the thermal management of low temperature devices. In this work, we investigate sub-1~K inter-chip thermal resistance of a flip-chip bonded assembly, where two silicon chips are interconnected by compression bonding via indium bumps. The total thermal contact area between the chips is 0.306~mm$^2$ and we find that the temperature dependence of the inter-chip thermal resistance follows the power law of $\alpha T^{-3}$, with $\alpha = 7.7-15.4$~K$^4$~\textmu m$^2$/nW. The $T^{-3}$ relation indicates phononic interfacial thermal resistance, which is supported by the vanishing contribution of the electrons to the thermal conduction, due to the superconducting interconnections. Such a thermal resistance value can introduce a thermalization bottleneck in particular at cryogenic temperatures. This can be detrimental for some applications, yet it can also be harnessed. We provide an example of both cases by estimating the parasitic overheating of a cryogenic flip-chip assembly operated under various heat loads as well as simulate the performance of solid-state junction microrefrigerators utilizing the observed thermal isolation. 

\end{abstract}

\maketitle

3D integration techniques enable vertical interconnection of different microsystems in order to enhance the overall system functionality. In general, the techniques provide higher packing density of microcircuits compared to the conventional 2D planar fabrication~\cite{3Dint}, and importantly, the possibility to interconnect non-monolithically several device layers (chips), which have monolithically incompatible fabrication processes. Recently, 3D integration of the microsystems operated at cryogenic temperatures has gained lots of attention, thanks to the intensive growth of the quantum technology field. For superconducting quantum processors with large qubit counts, 3D integration is a must~\cite{qubitflipchip1} and different realizations already exist~\cite{qubitflipchip2, qubitflipchip3}. Large spin-qubit quantum processors and the interfacing Cryo-CMOS peripherals naturally benefit from 3D integration~\cite{spinqubit}, as well as schemes utilizing 2-qubit coupling with superconducting resonators~\cite{spinqubit2}. Another superconducting component that calls for similar approaches is a vertical solid-state microrefrigerator proposed in Ref.~\cite{sci.adv}

3D integration can pose challenges to the thermal management of microsystems~\cite{3D_thermal2}, and very little is known about the inter-chip thermal resistance in such assemblies at low temperatures. Due to the steep thermal suppression of electron-phonon coupling and interfacial phonon-phonon transport, several cryogenic electronic applications require careful design of the device-to-bath thermal contact stack to prevent overheating even under modest dissipative loads. Large local non-equilibrium populations of phonons can be generated during the operation of microdevices, and these phonons must be efficiently conducted away from the generated hot spots. The heat must be eventually dumped to the thermal bath formed, for example, by the (metallic) sample holder at the cold head of a cryostat. The interfaces and constrictions introduced to such thermalization paths by 3D integration are bound to increase the overall device--to--bath thermal resistance. This can be very beneficial to applications in superconducting solid-state microrefrigerators, provided that the thermal resistance occurs close to the cooled interface.

\begin{figure}[h!]
\includegraphics[width = 8.5 cm]{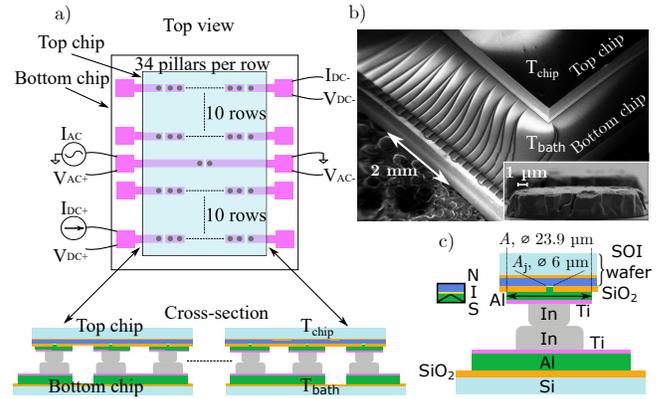}
\caption{\label{fig:device} a) Illustrative device schematic from the top and a single row cross-section with highlighted experimental wiring connections on the wire bonding pads. b) Scanning electron micrograph of the device and an indium bump prior bonding. c) The material stack of a single circular pillar: silicon--on--insulator (SOI), silicon dioxide (SiO$_2$), aluminum (Al), titanium (Ti), indium (In) and silicon (Si). Thermal contact area is defined by the circular aluminum collar with the diameter of 23.9~\textmu m and the tunnel junction area $A_\textrm{j}$ is similarly defined by the diameter of 6~\textmu m. The full device has a total of 682 interconnection pillars distributed in two arrays of 340 pillars adjacent to the pillar pair on the center of the device.}
\end{figure}

\vspace{1cm}
In this work, we report on sub-1 K thermal resistance characterization of a 3D integrated assembly, where two silicon chips are joined by indium bumps with flip-chip atmospheric compression bonding. We find that the thermal resistance of the device follows $T^{-3}$ power law, with several times larger absolute value than predicted by the diffusive mismatch theory. We simulate $T^4$ phononic power law to show how non-negligible heat loads can pose a thermalization problem in flip-chip devices operating at sub-1 K temperature range. We also provide a simulation of vanadium (V)- and aluminum (Al) based superconducting microrefrigerators using the measured thermal resistance and previously demonstrated junction parameters~\cite{APL_antti}.

The system analyzed in this letter is a flip-chip device consisting of a bottom chip of 8~x~8~mm$^2$ and a top chip of 4~x~6~mm$^2$ surface footprint. Figure~\ref{fig:device}a) shows an illustrative schematic of the flip-chip device from the top of the assembly and a cross-section of a single interconnection row. The device comprises a total of 682 interconnection pillars, with two arrays consisting of 340 pillars each, positioned adjacent to the pillar pair depicted at the center of the device. A scanning electron micrograph in Fig.~\ref{fig:device}b captures the device, alongside a single indium bump before the bonding process. The interconnection of chips is facilitated by indium bumps on both the bottom and top chips, forming a unified interconnection pillar as presented in Fig.~\ref{fig:device}c. Each of these pillars houses a normal metal--insulator--superconductor (NIS) junction, where the normal metal component consists of a pool of degenerately-doped silicon, common to all of the junctions. In this work, a junction pair in the middle of the device was used as a thermometer to measure the temperature of the top chip as a function of the applied heating power. The heating was applied through two junctions located on the diagonal corners of the device. The experimental connections are highlighted in Fig. \ref{fig:device}a) and will be discussed in more detail following the description of the fabrication process.

The bottom chip was fabricated on a 150~mm \textit{p}-type silicon (Si) wafer using thermal oxide growth (100~nm SiO$_2$), metal sputtering [200~nm Al and 50~nm titanium (Ti)], ultraviolet (UV) lithography and reactive ion etching. In bumps with thicknesses of 3~\textmu m were deposited with an electron-beam evaporator on a patterned 5~\textmu m thick lift-off resist. Prior to the evaporation, the Ti surface was argon-ion beam etched to remove the surface oxide \textit{in-situ}. After the evaporation, the resist was removed in an acetone bath, completing the bottom chip fabrication. The top chip was fabricated on a 150 mm silicon--on--insulator (SOI) wafer, in which the 40~\textmu m thick device layer has an \textit{n}-type (phosphorus) doping yielding an electrical resistivity less than 1.5 m$\Omega$cm at room temperature. Additional ion implantation of phosphorus was done with an acceleration voltage of 27 kV and a fluence of 4$\times $10$^{15}$ cm$^{-2}$ to increase the surface dopant concentration. Based on the thermal budget of the following process steps, this yields a total dopant concentration of about 10$^{20}$~cm$^{-3}$ for the surface layers. Even though the volumetric electron-phonon coupling in doped semiconductors is lower than that of metals~\cite{dopingMika}, the large volume of the normal metal ($4000\times 6000\times  40$~\textmu m$^3$) ensures strong overall coupling. A 100~nm layer of SiO$_2$ was grown on the top chip wafer, followed by the tunnel junction contact etching. The $A_\textrm{j} \approx 28$~\textmu m$^2$ sized contacts to the Si surface were done by UV lithography combined with plasma- and wet etching. The tunnel barrier was grown thermally by keeping the Si surface under 0.13~mbar of pure oxygen for 10 minutes at 550~$^\circ$C. A~500 nm Al layer and an additional 25~nm Ti layer were sputtered \textit{in-situ}. The Ti layer functions as an adhesion layer between Al and In, which was deposited, as described above, to complete the top chip fabrication.

We define the thermal interface area per pillar by the tunnel junction area and the surrounding Al collar as shown in Fig.~\ref{fig:device}c). With 682 pillars in the device, the total thermal interface area is $A = $~0.306~mm$^2$. Both wafers were diced to individual chips, with each wafer yielding from a hundred to several hundred devices. The bottom- and the top chips were connected together with a flip-chip bonder using a room temperature compression cycle. The cycle consisted of an alignment-, a chip parallelization- and a 30 second compression step with a force of 6.8~N at 29~$^\circ$C. With the given recipe, we have observed a 50\% reduction in the height of the bumps. Under the assumption that the total volume of the bumps remains constant, this transformation results in a twofold expansion of the surface area of the indium. Conversely, the surface area of the indium oxide remains unaltered, leading to an effective In-In contact. The method is discussed more thoroughly in the supplementary material of Ref.~\cite{qubitflipchip2}. The device was attached to the copper sample holder with a polyvinyl butyral based glue used commonly in experimental low temperature physics. The sample holder was mounted with metallic screws on the mixing chamber flange of a ${}^{3}\textrm{He} / {}^{4}\textrm{He}$ dilution refrigerator to ensure a good thermal contact from the cryostat to the sample holder and the bottom chip. The efficient thermalization allowed us to measure the temperature of the bottom chip $T_\mathrm{bath}$ with a calibrated ruthenium oxide thermometer positioned on the mixing chamber flange.

\begin{widetext}
    
\begin{figure*}[h!]
\centering
\includegraphics[width = 17.0 cm]{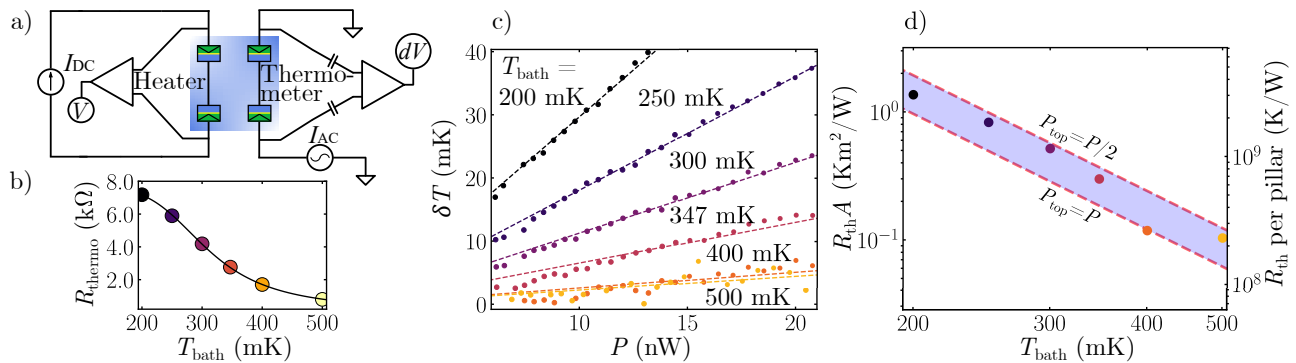}
\caption{\label{fig:maindata} a) Experimental schematic: the heater is DC current biased $I_\textrm{DC}$ with a floating supply and the corresponding voltage $V$ is read across the heater. The thermometer is biased with an AC current $I_\textrm{AC} = 0.5$~nA at 18 Hz and the response $dV$ is recorded with an AC coupled lock-in amplifier. b) Zero-bias response of the thermometer at 200~mK to 500~mK with a fitted tunneling model. c) Measured top chip temperature difference relative to the bath temperature $\delta T =~ T_\textrm{top} - T_\textrm{bath}$ as a function of the applied total power to the heater. d) Effective interfacial thermal resistance as a function of the bath temperature. Dashed lines represent the two heat distribution limits for Eq. (\ref{eq:alpha}) fitting. Data points correspond to 75\% heat injection to normal metal. Right side y-axis shows the thermal resistance per pillar as a supplementary figure of merit.}
\end{figure*}
\end{widetext}

The thermal resistance of the device was determined by heating the top chip with the heater junction pair and measuring the elevated temperature of the top chip $T_\textrm{top}$ with the thermometer junctions at various bath temperatures $T_\textrm{bath}$. Figure \ref{fig:maindata}a) supplements the experimental wiring provided in Fig. \ref{fig:device}a). The heater was biased with a floating DC current $I_{\textrm{DC}}$ and the corresponding voltage $V$ over the heater was recorded with a junction field-effect transistor amplifier and a digital multimeter. The thermometer was operated with a sinusoidal AC current of $I_{\textrm{AC}}=$~0.5~nA at 18~Hz and the voltage across $dV$ was recorded with an AC coupled lock-in amplifier. The thermometer was calibrated against the bath temperature by fitting the zero-bias resistance $R_\textrm{thermo}$ against $T_\textrm{bath}$ to a standard model used for NIS junctions \cite{coolerReview} and assuming that the electron- and the phonon temperatures are equal. The fit yielded a normal state tunneling resistance R$_{\textrm{T}} = $~107 $\Omega$, a sub gap leakage parameter $\gamma =$~1.4$\times $10$^{-2}$ and a critical temperature $T_\textrm{c} = $~1.12 K. Figure \ref{fig:maindata}b) shows the calibration with the data points and the fit line. We note that the self heating of the thermometer was smaller than 2.5~fW, thus negligible compared to the applied heater power values of the order of 10~nW.

Figure \ref{fig:maindata}c) shows the temperature increase of the top chip $\delta T = T_\textrm{top} - T_\textrm{bath}$ as a function of the total heating power $P$ of the heater junctions for $T_\textrm{bath}$ between 200~mK and 500~mK. The relation between $\delta T$ and $P$ is linear for $\delta T < 40$~mK and $P > 5$~nW, which corresponds to DC bias voltages above the superconducting energy gap, i.e. $V > 2\Delta/e$ for given junction parameters. In this linear regime, the thermal resistance $R_\textrm{th}$ between the bottom and the top chip at given $T_\textrm{bath}$ is
\begin{equation}
    R_\textrm{th} = \frac{\delta T}{P_\textrm{top}},
    \label{eq:gamma}
\end{equation}
where $P_\textrm{top}$ is the power injected to the top chip. For NIS junctions biased at $V \gg 2\Delta/e$, the heat injected into the normal metal approaches 50\% of the total heat generation. Given that a portion of the heat injected into the superconductor might also flow to the top chip due to insufficient quasiparticle evacuation from the vicinity of the junction, in the following we consider the effective heating power to the top chip in to lie in a bracket between the values $P_\textrm{top}\approx P$ and $P_\textrm{top}= P/2$.

Scaling the extracted thermal resistances by $A$ yields an effective interfacial thermal resistance (eITR), which follows the temperature dependence of \cite{Swartz_thermalInterface} 
\begin{equation}
R_\textrm{th}A = \alpha T^{-3}_\textrm{bath}.
\label{eq:alpha}
\end{equation}
Figure~\ref{fig:maindata}d) shows eITR as a shaded blue area defined by the two dashed red lines, obtained by fitting the constant $\alpha$ to the extracted values of $R_\textrm{th}A$ at different temperatures. The lower boundary is produced by the assumption of full heat dissipation to the top chip ($P_\textrm{top}= P$) and the upper boundary by partial dissipation ($P_\textrm{top}= P/2$). The data points correspond to the average of these limits, i.e. 75\% of $P$ is injected to the normal metal. The lower estimate yields the prefactor of $\alpha = $~7.7~K$^4$\textmu m$^2$/nW, which doubles for the upper line fit as $\alpha = $~15.4~K$^4$\textmu m$^2$/nW.

Since In is a good thermal conductor at low temperatures \cite{Graham2014} and the thicknesses of the rest of the layers along the pillars are smaller than the dominant thermal phonon wavelengths, it is unlikely to have any significant thermal gradients along the interconnecting columns. Hence, we assume the origin of the observed $R_{\mathrm{th}}$ to be of interfacial nature. Adhesion between surfaces is a complex problem involving various types of bonds\cite{Gerberich_2006}. Obtaining high quality metal-insulator interfaces is generally more challenging in comparison with metal-metal contacts due to the different properties of the connecting materials. This points the largest component of the observed eITR to the diffusive Al--SiO$_2$ interfaces at the tunnel junction and the surrounding collar. Due to their largely different oxide thicknesses (below 3~nm at the junction and 100~nm at the collar), these regions may isolate heat at different capacities, yet the two contributions cannot be assessed separately in our measurement. Practically, the effective thermal contact area is dominated by the collar, as its area is relatively large compared with $A_\mathrm{j}$ (1:15). The eITR of a metal-substrate interface can be quantitatively described in multiple systems by the acoustic mismatch (AMM) and diffusive mismatch (DMM) models \cite{Little_1959_AMM,Swartz_thermalInterface,Swartz_1989}, which differ minimally at low temperatures. For Al--SiO$_2$, DMM gives a value of $\alpha_\textrm{eff}=~1.1$~K$^4$\textmu m$^2$/nW, whereas our measurements indicate 7.1--14 times larger eITR. Previous works with metal films on oxidized Si substrates have shown up to five times larger eITR compared with DMM calculations~\cite{2007_Rajauria_phonon,2013_Pascal_phonon,Wang_2019_thermal_resistance} and a fourfold increase in eITR has been observed by depositing a thin adhesive Ti layer between a metal and a substrate \cite{Wang_2019_thermal_resistance}. Similarly, the multiple interfaces in the inter-chip pillars in our device may contribute non-trivially to additional thermal resistance. DMM predicts a value of 3.5~K$^4$\textmu m$^2$/nW to the full stack, as discussed more in the supplementary material. However, the applicability of the DMM model to the presented system with multiple interfaces is questionable, since modeling the full stack by adding individual interface contributions independently together might not fully capture phonon propagation in a stack with layers of thicknesses comparable to the dominant phonon wavelength.

\begin{widetext}

\begin{figure*}[ht!]
    \centering
    \includegraphics[width = 17cm]{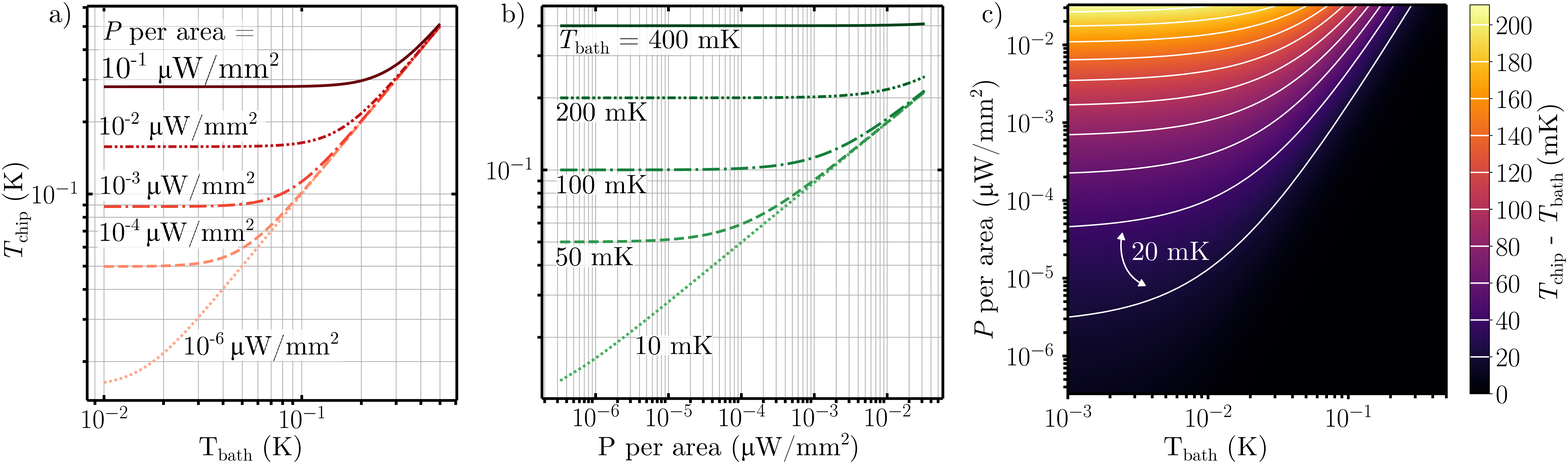}
    \caption{Simulation of $T^4$ phononic power law given in Eq. (\ref{eq:heatsimu}) using the eITR parameter $\alpha = $ 15.4 K$^4$\textmu m$^2$/nW. a) Temperature of the chip as a function of the bath temperature for power density values ranging from 10$^{-6}$ \textmu W/mm$^2$ to 10$^{-1}$ \textmu W/mm$^2$. b) Temperature of the chip as a function of the power density for bath temperature values between 10~mK and 400~mK. c) Colormap of the absolute overheating of the chip as a function of the bath temperature and the power density.}
    \label{fig:heatsimu}
\end{figure*}

\end{widetext}

Large eITR can be detrimental for applications requiring good thermal contact, and care must but taken in device design to mitigate potential thermal bottlenecks. However, the large eITR can also be a desired feature in some devices, such as solid-state microrefrigerators. Here, we present simulations supporting both claims. In the following cases, the eITR parameter is set to $\alpha = $ 15.4 K$^4$\textmu m$^2$/nW.
Figure \ref{fig:heatsimu} shows the thermal behaviour of a system operating at sub-1 K temperature range with various heat loads per thermal contact area ($P/A$), which is modelled by solving the phononic heat flow law
\begin{equation}
P/A = \frac{1}{4 \alpha} (T_\textrm{bath}^4 - T_\textrm{chip}^4),
\label{eq:heatsimu}
\end{equation}
where the chip $T_\mathrm{chip}$ and the thermal bath $T_\mathrm{bath}$ are decoupled through the thermal resistance characterized by $\alpha$. Equation~(\ref{eq:heatsimu}) was solved for $T_\mathrm{chip}$ varying the power density $P/A$ and the bath temperature $T_\mathrm{bath}$. Figure \ref{fig:heatsimu}a) shows the chip temperature as a function of the bath temperature for few values of power density ($P$ per area). When scaled with the thermal interface area $A$ of the device analyzed in this letter, the power values range from around 300~fW to 30~nW. At 10~mK, the usual base temperature of a commercial dilution refrigerator, increasing power density corresponds to an elevated chip temperature $T_\mathrm{chip}$ of 16~mK at 10$^{-6}$~\textmu W/mm$^2$ and notably 280~mK at 10$^{-1}$~\textmu W/mm$^2$. At higher bath temperatures the chip temperature coincide with the bath temperature, i.e. $T_\mathrm{chip} = T_\mathrm{bath}$. Figure~\ref{fig:heatsimu}b) shows the temperature of the chip as a function of power density for few values of bath temperature. Again, the chip temperature is elevated significantly at $T_\mathrm{bath} = $~10~mK with increasing power density. At intermediate temperatures the chip is effectively thermalized to the bath at the lowest power density values and at $T_\mathrm{bath} = $ 400~mK the elevated $T_\mathrm{chip}$ is negligible even at highest power density values. The overall behaviour is summarized in Fig. \ref{fig:heatsimu}c), where a colormap shows the absolute overheating of $T_\mathrm{chip}$ from the bath as a function of the two variables, bath temperature and power density. In conclusion, the simulation exemplifies to what extent the flip-chip stack can act as a thermalization bottleneck at power density values relevant to the devices operating at sub-1 K temperature range. To overcome this, the straightforward approach is to increase the thermal interface area i.e. the diameter and/or amount of the interconnect pillars, when applicable to the device design. However, practical limitations may arise when attempting to optimize interconnect density using the current indium bump technology e.g. due to bonding and fabrication related limits. Although alternative bonding methods, such as Cu-Cu hybrid bonding\cite{hybridbonding}, have been developed to enhance interconnect density, there is currently no superconducting interconnect available that offers both low bonding force and temperature, but new methods are actively being developed. We note that normal metal interconnects exhibit a thermal conductivity that encompasses not only phononic but also an electronic contribution, which allows better thermalization for technologies with non-superconducting interconnects.

In contrast to the thermalization bottleneck, the large eITR can be harnessed in optimizing superconducting solid-state microrefrigerators. Here we present a simulation of V- and Al based superconducting microrefrigerators by using the eITR prefactor $\alpha$ obtained in this work and the junction parameters obtained in our previous work \cite{APL_antti}. In the simulation, we assume a strong electron-phonon coupling in the normal metal, such that the electron- and the phonon systems are at an equilibrium temperature i.e. $T_\textrm{N,e} = T_\textrm{N,ph} = T_\textrm{N} $. We also assume that the superconductor is thermalized to the bath temperature, i.e. $T_\textrm{S,e} = T_\textrm{bath}$. The simulation parameters are shown in Tab. \ref{tab:params} and the heat balance is given by

\begin{equation} \label{eq:simulation}
\begin{split}
&\dot Q_{\textrm{NIS}} + \dot Q_{\textrm{ph-ph}} = 0, \textrm{where} \\
&\dot Q_{\textrm{NIS}} = \frac{A_\textrm{j}}{e^2R_{\textrm{A}}} \int_{-\infty}^{\infty}\textrm{d}E (E - eV_\mathrm{b})\rho_\textrm{S}(E)\\ & \qquad  \times [f_\textrm{S}(E,T_\textrm{bath}) - f_\textrm{N}(E - eV_\mathrm{b}, T_\textrm{N})] \\
&\dot Q_{\textrm{ph-ph}} = \frac{A}{4 \alpha} (T_\textrm{bath}^4 - T_\textrm{N}^4), \\
\end{split}
\end{equation}

where $\dot Q_\textrm{NIS}$ is the electronic heat flow to the normal metal, $\dot Q_\textrm{ph-ph}$ is the phononic heat flow to the normal metal from the bath, $E$ is the energy, $V_\mathrm{b}$ is the voltage bias, $\rho_\textrm{S} = \left| \mathbb{R} \left(\frac{E / \Delta + i\gamma}{\sqrt{(E / \Delta+i\gamma)^2 - 1}} \right) \right|$ is the normalized Dynes density of states of the superconductor and $f_\textrm{S}$ $(f_\textrm{N})$ is the Fermi-Dirac distribution of the superconductor (normal metal).

\vspace{1cm}

The performance of V- and Al based coolers was estimated by solving $T_\textrm{N}$ from the heat balance equation (Eq.~(\ref{eq:simulation})) as a function of $V_\mathrm{b}$ for devices with two different ratios between $A$ and $A_\textrm{j}$. The solid lines in Fig. \ref{eq:simulation} represent a device with an Al collar placed around the junction ($A = 16\cdot A_\textrm{j}$), similar to the structure studied in this letter. The dashed lines show approach, where the collar is left out and the heat leakage occurs entirely through the junction ($A = A_\textrm{j}$). Here it is assumed, that the eITR is equal at the junction and the collar. The V stage is able to cool down from the bath temperature of 890~mK to 740~mK with the collar structure, and to 350~mK in the approach without collar. The performance of the Al stage is not affected as much due to the $T^{4}$ dependent heat leak. The Al stage can cool down from 350~mK to 170~mK with the collar, and to 131~mK without it. The relative cooling with V from $T_\textrm{bath}$ is 60\% at best. We note that the Al stage can, in principle, be thermalized to the V stage to form a two-stage microrefrigerator. More quantitative modeling and discussion of a multi-stage cascaded microrefrigerator can be found in Ref. \cite{APL_antti} including the role of transparency-limited junction leakage.

\begin{table}[h!]
\caption{The simulation parameters for the V- and Al based microrefrigerators. The effective interfacial thermal resistance fit prefactor $\alpha$ is common for both stages, presented in this letter. The specific electrical resistance for the tunnel junctions $R_\textrm{A}$, the sub gap leakage parameter $\gamma$ and the superconducting energy gap $\Delta/e$ are our recently demonstrated parameters \cite{APL_antti}.}
\begin{tabular}{ccccc} \toprule
\hline \rule{0pt}{3ex} 
          Superconductor & $\alpha$ &$R_\mathrm{A}$                                              & $\Delta / e$                   & $\gamma$                                   \\
         & (K$^4$\textmu m$^2$/nW) & ($\Omega$\textmu m$^2$) &  (\textmu$V$)  &  \\\hline \rule{0pt}{3ex} 
V & 15.4 & 71   & 540  & 6$\times $10$^{-3}$  \\
Al & 15.4 & 48   & 182  & 8$\times $10$^{-3}$
\end{tabular}

\label{tab:params}
\end{table}

In summary, we have measured the thermal resistance of a superconducting flip-chip assembly at sub-1~K temperatures and compared the result to the DMM theory and previous experiments. Our results exemplify a case in which advanced integration of superconducting assemblies would result in low thermal contact between the chips in 3D integrated assembly. We showed how the thermal resistance can pose a bottleneck for the thermal management of such a system at sub-1~K temperatures due to the demonstrated thermal resistance. We also presented a simulation of the utilization of the measured thermal resistance for the purpose of thermal isolation in V and Al based superconducting solid-state microrefrigerators.

\begin{figure}[h!]
\includegraphics[width = 7.5 cm]{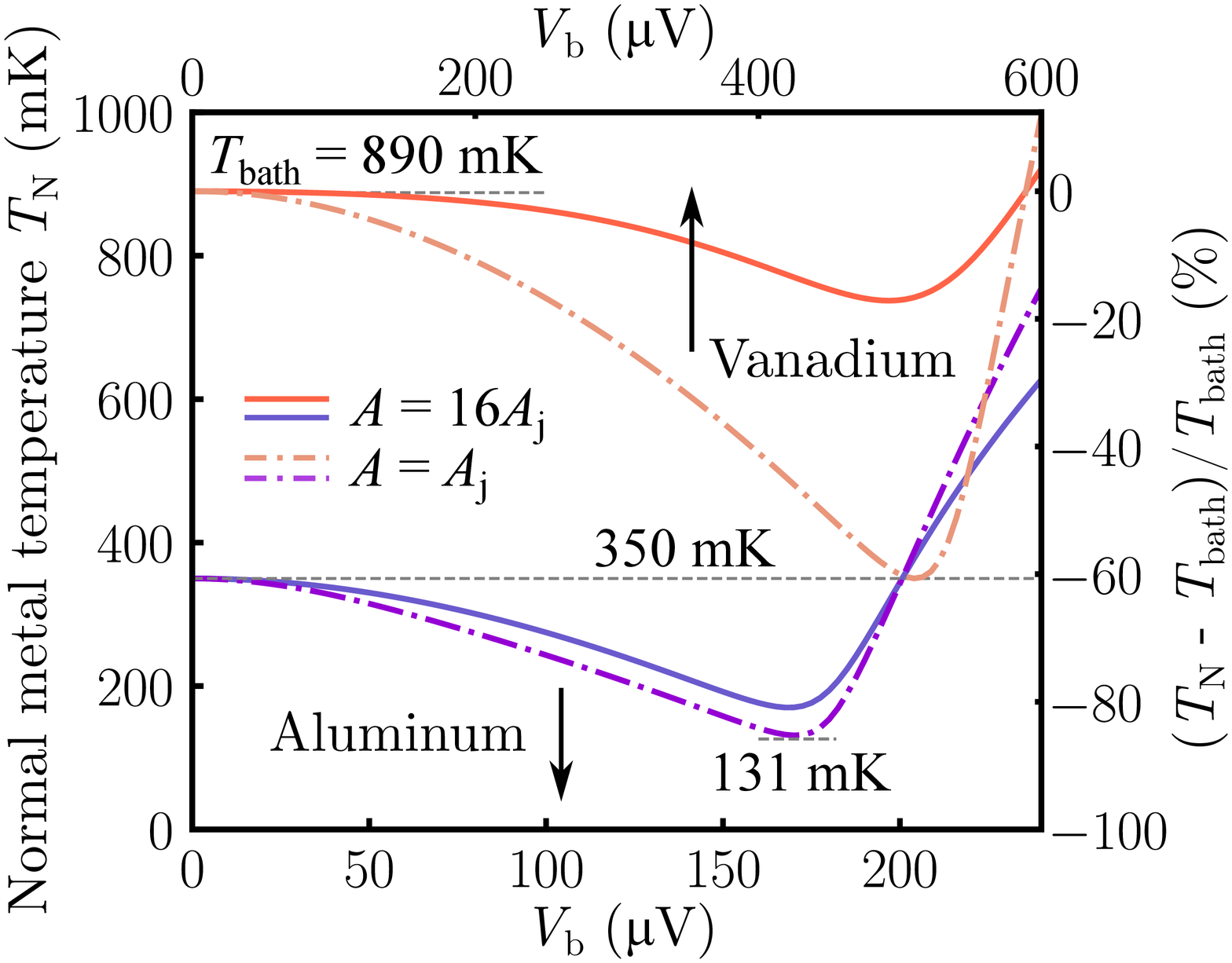}
\caption{\label{fig:coolingSimulatio} Simulated cooling performance of V- and Al based microrefrigerators as a function of voltage bias $V_\textrm{b}$. Solid lines represent the structure presented in this letter, with the collar dominating the thermal interface area over the junction. Dashed lines show the optimal configuration, where the thermal interface area is equal to the junction area. The simulation parameters are given in Tab.~\ref{tab:params} and the simulation model is given in Eq.~(\ref{eq:simulation}). The normal metal temperature $T_\textrm{N}$ represents both the electron- and the phonon temperature due to the strong electron-phonon coupling.}
\end{figure}

\section*{Supplementary Material}
The models and parameters used for calculating the diffusive mismatch model predictions for the Al--SiO$_2$ and full material stack are provided as a supplementary material.

\section*{Acknowledgements}
The authors thank T. Häkkinen, J. Toivonen, L. Grönberg and J. Salonen for technical assistance in the sample fabrication in VTT and OtaNano Micronova cleanroom facilities. We also thank L. Wang for discussions. The research was funded by the European Union's Horizon 2020 research and innovation programme under grant agreement No.~766853 EFINED, No.~824109 European Microkelvin Platform (EMP), No.~101113086 SoCool and No. 101007322 ECSEL Joint Undertaking (JU), the Academy of Finland through project No.~322580 ETHEC and the QTF Centre of Excellence project No.~336817, Business Finland through Quantum Technology Industrial (QuTI) No.~128291. We also acknowledge Technology Industries of Finland Centennial Foundation for funding.

\section*{Author declarations}
J. Hätinen contributed to process development, characterization, data analysis and made the first version of the manuscript. E. Mykkänen contributed to characterization, data analysis and writing. K. Viisanen made the theoretical comparison and contributed to the writing. A. Ronzani contributed to the conceptualization, characterization and writing. A. Kemppinen contributed to the conceptualization and writing. L. Lehtisyrjä contributed to the process development and writing. J. S. Lehtinen conceptualized and lead the work, contributed to the process development, characterization and writing. M. Prunnila conceptualized and supervised the work, and contributed to the writing.

The authors have no conflicts to disclose.

\section*{Data availability}
The data that support the findings of this study are available from the corresponding author upon reasonable request.

\section*{References}
\bibliography{main}

\end{document}